\documentstyle[prd,aps,epsf,twocolumn]{revtex}

\begin{document}
\draft
\preprint{DUKE-TH-98-161}

\title{Lattice QCD with Ginsparg-Wilson fermions\cite{DOE}}
\author{Shailesh Chandrasekharan\cite{EMAIL}}
\address{ 
Department of Physics, Box 90305, Duke University, \\
Durham, North Carolina 27708, USA }

\date{September 20, 1998}
\twocolumn[\hsize\textwidth\columnwidth\hsize\csname @twocolumnfalse\endcsname
\maketitle

\begin{abstract}
Lattice QCD using fermions whose Dirac operator obeys the Ginsparg-Wilson 
relation, is perhaps the best known formulation of QCD with a finite 
cutoff. It reproduces all the low energy QCD phenomenology associated 
with chiral symmetry at finite lattice spacings. In particular it 
explains the origin of massless pions due to spontaneous chiral symmetry 
breaking and leads to new ways to approach the $U(1)$ problem on the 
lattice. Here we show these results in the path integral formulation and 
derive for the first time in lattice QCD a known formal continuum relation
between the chiral condensate and the topological susceptibility.
This relation leads to predictions for the critical behavior of the 
topological susceptibility near the phase transition and can now be checked 
in Monte-Carlo simulations even at finite lattice spacings. 
\end{abstract} 
]
%



\section{INTRODUCTION}

  Formulating QCD non-perturbatively is important for
understanding the low energy phenomenology of hadronic physics. 
The only such formulation that is known at present is through the 
lattice regularization. However, until now fermions have been a major 
obstacle. The main reason has been that 
chiral symmetry is easily broken on the lattice. Since two of 
the main features of low energy QCD, namely the spontaneous breaking
of chiral symmetry and the anomaly, are intimately connected with
chiral symmetry, one hardly uses the lattice discretization to
discuss the low energy phenomenology of QCD. Even if such a discussion
was carried out, one usually needs to invoke continuum limits 
before the formulation can actually be argued to reproduce all 
the low energy properties of QCD.

At finite lattice spacings it is usually unclear as to how chiral 
symmetry is realized and what properties of QCD are lost. For example 
with Wilson fermions\cite{Wilson}, 
since chiral symmetry is explicitly broken,
there is no reason to expect massless pions unless one tunes a mass 
parameter to a critical point. Further, the critical point appears 
to describe spontaneous 
breaking of parity and flavor symmetry\cite{Aoki}. Thus the soft
pion theorems of QCD will not be reproduced at finite lattice spacings.
If instead one starts with staggered fermions\cite{Stagg}
one can only formulate four flavors of QCD in the Lagrangian formulation. 
In this case a $U(1)$ subgroup of the flavor
non-singlet chiral symmetry is exact on the lattice. However,
the flavor symmetry gets broken at finite lattice spacings which
again is expected to be restored only in the continuum limit.

The anomalous breaking of the flavor singlet chiral symmetry is expected 
to give the $\eta^\prime$ particle its mass. With both Wilson and staggered 
fermions, discussions involving the anomaly have new complications. Since 
the lattice regularization breaks the anomalous chiral symmetry 
along with other chiral symmetries, the effects of the anomaly cannot
be easily isolated. The simple continuum discussion in terms of the zero 
modes of the Dirac operator and topology of the gauge fields breaks down.
All this indicates that both Wilson and staggered fermion formulations are 
not ideal for a description of low energy QCD phenomenology at finite 
lattice spacings.

  A long time ago Ginsparg and Wilson\cite{Ginsparg} showed that 
if the Dirac operator, $D$, obeyed the relation
\begin{equation}
\label{eq:GWR}
D\gamma_5 + \gamma_5 D = a D \gamma_5 D,
\end{equation}
where $a$ is the lattice spacing,
then the theory would have a remnant chiral symmetry up to contact
terms. An year ago Hasenfratz noticed that the fixed point action of
QCD obeyed the Ginsparg-Wilson relation\cite{Hasenfratz0} and hence
a series of interesting results\cite{Hasenfratz1,Hasenfratz2} about
renormalization and lattice index theorems followed. The new massless
fermion formulations motivated by the overlap\cite{Neuberger} also obeyed 
the Ginsparg Wilson relation. It seemed that there was more to the relation 
than was appreciated in the past. Recently L\"{u}scher showed that there 
is an exact chiral symmetry on the lattice\cite{Luscher} due to the 
Ginsparg-Wilson relation, since the Nielsen-Ninomiya theorem\cite{Nielsen} 
is not applicable. As a result, for the first time, we can write down a 
formulation of QCD with the right internal symmetries of the theory at 
finite lattice spacings. Thus in this formulation the role of the continuum 
limit is merely to renormalize some multiplicative constants and suppress 
finite lattice artifacts which do not change the internal symmetry structure 
of the theory. This leads to the most elegant non-perturbative formulation 
of QCD.

  In this paper we review the main features of the above observations. We
use the exact chiral symmetry that arises due to the Ginsparg-Wilson 
relation to derive the essential physics of low energy QCD in the
path integral formulation. For the first time we are able to discuss the 
physics of chiral symmetry breaking relevant to QCD with any number of 
flavors, the associated pion physics and the essential ingredients of 
the $U(1)$ problem, all directly at finite lattice spacings without much 
effort. In particular we derive a chiral condensate of the theory which 
reflects the breaking of chiral symmetry. For two or more flavors we 
reproduce the Goldstone's theorem if the chiral symmetry is spontaneously 
broken. Further, since the Ginsparg-Wilson relation allows for the breaking 
of the anomalous chiral symmetry through exact zero modes of the Dirac 
operator like in the continuum, we can easily discuss a solution to the 
$U(1)$ problem. By introducing an explicit quark mass we can 
derive a well known continuum relation, between the chiral 
condensate and the topological susceptibility, in lattice QCD.
The problems associated with the quenched approximation can be traced 
to the violation of this relation. Near the chiral phase transition 
at finite temperatures, this relation also leads to predictions for the 
critical behavior of the topological susceptibility which have not 
been appreciated before. These predictions can be verified in Monte-Carlo 
simulations.

Most of the results that we derive already appear in 
\cite{Hasenfratz1,Hasenfratz2,Neuberger2}. However, we have included them
for the coherent presentation of our work, and as far as we know this 
is the first time the results have been derived in the path integral 
formulation based on the symmetry discovered by L\"{u}scher\cite{Luscher}. 
This makes the re-derivations sufficiently different and perhaps more 
transparent. Our presentation of the $U(1)$ problem and its consequences 
are new. We have tried to keep the derivations simple and easily accessible 
to readers familiar with field theory. 

\section{CHIRAL SYMMETRY}
Let us begin by reviewing the exact lattice chiral symmetry, discovered 
by L\"uscher\cite{Luscher}, that arises due to the Ginsparg-Wilson relation.
The partition function of lattice QCD is given by
\begin{equation}
Z = \int [dU] {\rm e}^{-\frac{1}{g^2} S_G[U]} \int [d\bar\Psi] [d\Psi] 
{\rm e}^{(-S_F[\bar\Psi,\Psi])},
\end{equation}
where $S_G[U]$ is some lattice gauge action and $g^2$ is the bare gauge
coupling. The fermionic action is given by
\begin{equation}
S_F[\bar\Psi,\Psi] = - a^4\sum_{x,y} \bar\Psi(x)D_{x,y}\Psi(y),
\end{equation}
where we assume that $D$ obeys eq.(\ref{eq:GWR}).
We have suppressed color and flavor indices for convenience. 
The action is invariant under 
the infinitesimal chiral transformations of the form
\begin{equation}
\label{eq:chit}
\delta\Psi = \epsilon\gamma_5(1-a\frac{1}{2}D)\Psi,\;
\delta\bar\Psi = \epsilon\bar\Psi(1-a\frac{1}{2}D)\gamma_5.
\end{equation}
However, the measure is not invariant under these transformations.
In fact if $([d\bar\Psi] [d\Psi]) \equiv [d\phi]$
\begin{equation}
\label{eq:anom}
\delta(d[\phi]) = \epsilon a {\rm tr}(\gamma_5 D) [d\phi]
= -2\epsilon N_f (n_+-n_-) [d\phi],
\end{equation}
where $N_f$ is the number of quark flavors\cite{Luscher}. 
This means that the
flavor-singlet chiral transformation suffers an anomaly on the lattice
like in the continuum. Further like in the continuum the anomaly
arises only due to a non-zero index of $D$ which is given by $(n_+-n_-)$,
where $n_\pm$ are the number of zero modes of $D$ which are also 
simultaneous eigenstates of $\gamma_5$ with eigenvalue $\pm 1$.
We will explain the origin of eq.(\ref{eq:anom}) later. 
The flavored chiral transformations are obtained in the usual way by
including a traceless flavor matrix in the transformations
of eq.(\ref{eq:chit}). The flavored transformations are exact symmetries 
of the theory. Thus chiral symmetry and the anomaly emerge exactly like 
in the continuum, however all the above arguments are being made on a 
finite lattice.

A natural candidate for the chiral condensate emerges like in
the continuum. Starting from the definition of the (unnormalized) 
expectation value, $\langle \bar\Psi\gamma_5\Psi \rangle$ and performing 
a change of integration variables by an infinitesimal chiral 
transformation, it is easy to show that
\begin{equation}
\langle \bar\Psi\gamma_5\Psi\rangle = \langle \bar\Psi\gamma_5\Psi\rangle 
+ 2\epsilon \langle \bar\Psi (1-\frac{1}{2}aD) \Psi\rangle. 
\end{equation}
This means that if the theory has an exact chiral symmetry then
\begin{equation}
\langle \bar\Psi (1-\frac{1}{2}aD) \Psi\rangle = 0.
\end{equation}
When the theory contains more than one flavor, the above relation strictly 
holds at finite volumes. This is because there exist flavored 
(chiral) symmetries that are not broken by any anomalies. The only way 
the chiral condensate could be non-zero, is due to spontaneous breaking of 
chiral symmetry which is a subtle feature of the thermodynamic limit. To 
see this one has to add a small mass term to the Dirac operator and
break the symmetry explicitly. If the symmetry is spontaneously broken 
then the condensate would be non-zero when the thermodynamic limit is taken
before the explicit symmetry breaking mass is set to zero. In the case of 
a single flavor, the only chiral symmetry of the action is broken by the 
measure and $\langle \bar\Psi (1-\frac{1}{2}aD) \Psi\rangle \neq 0$ even 
in finite volumes. 

\section{PIONS AND THE U(1) PROBLEM}
In the presence of a mass term we can derive the modified relation
\begin{eqnarray}
\label{eq:mGWR}
&&(1+am)[(D+m)\gamma_5 + \gamma_5(D+m)]
\nonumber \\
&&= m(2+am)\gamma_5 +
a(D+m)\gamma_5(D+m),
\end{eqnarray}
using the Ginsparg-Wilson relation. To see the origin of massless
pions when the chiral symmetry is spontaneously broken, let us
look at the zero momentum pion correlation function $G_{ab}(p=0)$,
which is given by
\begin{equation}
G_{ab} \;=\; \frac{1}{\Omega}
\sum_{x,y}\langle \bar\Psi\;i\tau^a\gamma_5 \Psi(x)\bar\Psi\;i\tau^b\gamma_5 
\Psi(y)\rangle,
\end{equation}
where $\Omega a^4$ is the space-time volume. 
$\tau^{a,b}$ are the $N_f^2-1$ hermitian, traceless flavor matrices. 
Doing the Grassmann integral the above correlator can be rewritten as
\begin{eqnarray}
G_{ab} &=& 
\frac{1}{\Omega a^8} \left\langle {\rm tr} \left(
\frac{1}{D+m}\tau^a\gamma_5\frac{1}{D+m}\tau^b\gamma_5\right) 
\right\rangle
\nonumber \\
&=& \delta_{ab}\frac{1}{\Omega a^8} \left\langle {\rm tr} \left(
\frac{1}{D+m}\gamma_5\frac{1}{D+m}\gamma_5\right) 
 \right\rangle,
\end{eqnarray}
where we have absorbed the fermion determinant in the path integral
measure along with the Boltzmann weight of the gauge fields used to
calculate the expectation value.
By multiplying eq.(\ref{eq:mGWR}) with $(D+m)^{-1}$ from left and 
$(D+m)^{-1}\gamma_5$ from the right we get
\begin{eqnarray}
\lefteqn{{\rm tr} \left(
\frac{1}{D+m}\gamma_5\frac{1}{D+m}\gamma_5\right)\;=\; }
\nonumber \\
& &\frac{2(1+am)}{m(2+am)}
{\rm tr} \left(\frac{1}{D+m} - a\frac{1}{2(1+am)}\right),
\end{eqnarray}
which can be used to rewrite $G_{ab}$ as an expectation value of the
chiral condensate. In the chiral limit this leads to
\begin{equation}
\lim_{m\rightarrow 0} G_{ab} = \delta_{ab}
\frac{1}{m a^4} \langle \bar\Psi[1- \frac{1}{2}aD]\Psi \rangle.
\end{equation}
If chiral symmetry is spontaneously broken the right hand side
is singular in the chiral limit. This singularity shows that 
$m_\pi^2 \propto m$, since in the chiral limit we expect 
$G_{ab} \propto (m_\pi^2)^{-1}$. This is the Goldstone's theorem.

  It is straight forward to obtain a solution to the 
$U(1)$ problem in the above language. Consider the zero momentum 
flavor-singlet pion correlator
\begin{equation}
G_{00} \;=\; \frac{1}{\Omega}
\sum_{x,y}\langle \bar\Psi\;i\gamma_5 \Psi(x)\bar\Psi\;i\gamma_5 
\Psi(y)\rangle.
\end{equation}
The Grassmann integral now yields two terms
\begin{eqnarray}
G_{00} &=&
\frac{1}{\Omega a^8} \left\langle \left[{\rm tr} \left(
\frac{1}{D+m}\gamma_5\frac{1}{D+m}\gamma_5\right) \right.\right.
\nonumber \\
&& \left.\left. - 
\left\{{\rm tr} \left(\frac{1}{D+m}\gamma_5\right)\right\}^2\right]
\right\rangle,
\end{eqnarray}
where again we have absorbed the fermion determinant into the
Boltzmann weight used to calculate the expectation value.
The first term is familiar from the non-singlet pion correlator. 
The second term is non-zero only in configurations that have a 
non-zero index of the lattice Dirac operator, exactly like in the
continuum. This is because we can use eq.(\ref{eq:mGWR}) and 
eq.(\ref{eq:anom}) to show that
\begin{eqnarray}
{\rm tr}\left( \frac{1}{D+m}\gamma_5\right) = 
\frac{2 N_f}{m(2+am)} (n_+-n_-)
\end{eqnarray}
is even on a finite lattice. Thus we get
\begin{eqnarray}
\label{scorr}
G_{00} &=& \frac{(1+am)}{m(1+am/2)}
\langle \bar\Psi[1- \frac{a}{2(1+am)}(D+m)]\Psi \rangle
\nonumber \\
&&- \frac{(2 N_f)^2}{m^2(2+am)^2}\; \frac{1}{\Omega}
\langle (n_+-n_-)^2 \rangle.
\end{eqnarray}
This relation can be used to obtain new insight on the $U(1)$ problem.
A solution to the $U(1)$ problem requires $G_{00}$ to remain non-singular in
the chiral limit. It is clear that the first term on the right hand side 
is singular in the chiral limit, if 
$\langle \bar\Psi[1- \frac{1}{2}aD]\Psi \rangle\neq 0$.
This is expected to be true in finite volumes for one flavor QCD
due to the anomaly, and infinite volumes for two or more flavor QCD 
due to spontaneous chiral symmetry breaking. In both these cases, 
if $G_{00}$ must remain non-singular, the singular piece in the first term
must cancel with a corresponding piece in the second term. 
This requires
\begin{equation}
\label{eq:U(1)}
\frac{1}{\Omega}\langle (n_+-n_-)^2 \rangle 
= \frac{m}{N_f^2}\langle \bar\Psi[1- \frac{1}{2}aD]\Psi \rangle,
\end{equation}
at the leading order in the chiral limit. Thus we see that the solution 
to the $U(1)$ problem requires a connection between the lattice 
topological susceptibility defined using the index of the lattice Dirac 
operator, and the chiral condensate in the chiral limit. Such a connection
is known in the literature in the context of effective models\cite{Leutwyler} 
and the large $N_c$ limit\cite{Vecchia}. However, now the same result emerges
directly from lattice QCD.

At finite volumes in the one flavor case, eq.(\ref{eq:U(1)}) is
easy to understand since only sectors with unit topological charge 
contribute to both sides of the equation. However, if one takes 
the infinite volume limit before taking the chiral limit, the
relation is somewhat surprising. In this limit, the chiral condensate
is expected to get contributions from a non-zero density of small 
eigenvalues of the Dirac operator. This is the content of the Banks-Casher 
formula\cite{Banks} and will be derived later in the present context. The 
exact zero modes of the Dirac operator on the other hand are expected to be 
suppressed with the volume. Thus it is unclear why the topological
susceptibility, which is related to exact zero modes on the lattice,
is also related to the density of approximate zero modes as
eq.(\ref{eq:U(1)}) requires. Another interesting consequence of the equation 
is that the topological susceptibility goes linearly to zero with the mass,
independent of the number of flavors, if chiral symmetry is spontaneously 
broken. This appears to be a consequence of the fact that the thermodynamic 
limit and the chiral limit do not commute\cite{Leutwyler}.

The relation described by eq.(\ref{eq:U(1)}) is valid only in the
full theory of QCD where fermions can be created and destroyed
in the vacuum. In the quenched limit the left hand side is independent of 
the mass and number of fermion flavors, which only enter through the 
fermion determinant. Thus in 
the quenched theory the flavor-singlet pion correlator will show 
divergences in the chiral limit, since the singular pieces between the
two terms in eq.(\ref{scorr}) do not cancel. The form of these 
singularities have been predicted using chiral perturbation 
theory\cite{Sharpe} and can now be tested even at finite lattice 
spacings. Further, since exact zero modes of the 
Dirac operator are necessary for the anomaly, they are physical and occur 
even at finite lattice spacings. These produce new singularities in the 
quenched theory at finite volumes. In particular the chiral condensate 
is expected to diverge in the chiral limit due to the zero modes.

\section{CRITICAL BEHAVIOR OF ZERO MODES}

We can extend the above analysis to finite temperatures. 
It is clear that eq.(\ref{eq:U(1)}) will be valid at non-zero 
temperatures in the phase where chiral symmetry remains broken 
and if the flavor singlet pseudo-scalar correlator is free of 
infra-red divergences. It is usually believed that for two 
massless flavors, infra-red divergences arise at the chiral 
transition $T_c$, only due to diverging correlation lengths in the 
pseudo-scalar isovector(the pions) channel and the scalar isocalar 
channel(the sigma). This leads to the $O(4)$ critical behavior which is 
second order\cite{Wilczek}. If this is true, the flavor singlet correlator 
given in eq.(\ref{scorr}), must be free of infra-red singularities 
even at $T_c$. The chiral condensate is the order parameter of the phase 
transition and at $T_c$ has a critical behavior given by
\begin{equation}
\langle \bar\Psi[1- \frac{1}{2}aD]\Psi \rangle \sim m^{\frac{1}{\delta}}
\end{equation}
where $\delta=4.82\pm0.05$\cite{Baker} is an $O(4)$ critical exponent. 
The cancellation of the singular pieces in the right hand side of 
eq.(\ref{scorr}) leads to
\begin{equation}
\label{eq:crit}
\frac{1}{\Omega}\langle (n_+-n_-)^2 \rangle \sim m^{1+\frac{1}{\delta}}
\end{equation}
at $T_c$. This critical behavior of the topological susceptibility has
been ignored in the literature and can form a useful consistency check
of $O(4)$ critical behavior.

For $T > T_c$, the topological susceptibility need not be related to the 
chiral condensate since in this phase both terms in the eq.(\ref{scorr}) 
are finite in the chiral limit. However, if we assume that the 
thermodynamic and the chiral limits commute in the chirally symmetric phase, 
then for two or more flavors we can also predict
\begin{equation}
\frac{1}{\Omega}\langle (n_+-n_-)^2 \rangle \sim 
\left\{ \begin{array}{cc}
m & {\rm T < T_c} \cr
m^{N_f} & {\rm T > T_c} \cr \end{array}
\right.
\end{equation}
The prediction for $T > T_c$ is easily derivable from general 
considerations\cite{Sch2}. This behavior is consistent with what we know 
about the phase transition. For example, with three or more flavors the 
transition is expected to be first order. This is reflected in the dramatic 
change in the small mass dependence of the topological susceptibility. The
change in the two flavor theory is smaller and matches with the prediction
of eq.(\ref{eq:crit}) for a second order transition.

It is useful to look for the above critical behavior of the topological 
susceptibility in any study of the QCD
phase transition, since it reflects the interesting interplay between the 
chiral symmetry and the $U(1)$ anomaly. It has been appreciated that this 
interplay is important in determining the order of the phase transition 
for two light flavors\cite{Pisarski}. In particular if the anomalous $U(1)$ 
is small for some dynamical reason, the fluctuations from new degrees of 
freedom can make the transition first order. On the other hand if the 
symmetry is broken by a large amount then the phase transition is expected 
to be in the $O(4)$ universality class. 

The fermionic definition of the topological susceptibility, used in the above
analysis can in principle be measured on the lattice since it involves 
calculating zero modes of the Dirac operator in a realistic simulation. 
In the staggered fermion formulation, which is most suited for studying 
the chiral phase transition, zero modes are shifted due to finite lattice 
spacing effects\cite{Smit}. Such effects distort the critical behavior of 
the zero modes, and one needs to invoke the continuum limit to see the 
effects of the anomalous symmetry emerge at the QCD phase 
transition\cite{Kogut,Sch2}. However, approaching the continuum limit is
quite difficult. On the contrary, there is evidence\cite{Edwards} that 
with Ginsparg-Wilson fermions the continuum limit is not necessary to see 
the zero modes that are responsible for the anomalous symmetry breaking. Thus 
the above critical behavior should emerge even at finite lattice spacings.

\section{DISCUSSION AND CONCLUSIONS}
All the above results have been derived without specifying an explicit 
realization of the the Dirac operator that obeys the Ginsparg-Wilson
relation. However, it is important that the Dirac operator does not
suffer from doubling, since it is possible to find Dirac operators which
obey the Ginsparg-Wilson relation but do not solve the fermion doubling
problem. In such cases, the real dynamics of QCD will not be reproduced.
The chiral symmetry discussed could be flavored as in staggered fermions.
The two known classes of solutions based on the fixed point 
actions\cite{Hasenfratz1,Uwe} and the overlap formulation\cite{Neuberger}
are expected to solve the doubling problem completely. The proposal by
Neuberger has a simple structure based on the four dimensional Wilson 
like Dirac operator $D_W$, and is given by 
$aD = 1 + D_W(D_W^\dagger D_W)^{-1/2}$. The important point is that $D_W$ 
must have an appropriately tuned negative mass term
to be in the right phase to produce QCD.   Whether such a phase exists 
is currently being explored. A further effect of the Wilson mass is to 
renormalize the theory and control leading lattice artifacts.

The Neuberger Dirac operator is also closely related to the Shamir's 
variant of the domain wall fermions\cite{Shamir}, when the distance 
between the walls is taken to infinity. It is likely that all the nice
properties discussed above will emerge with small violations when the
walls are separated by a large but finite separation. This strategy can
be used for practical simulations and still retain the interesting 
properties of Ginsparg-Wilson fermions. Preliminary tests for QCD has 
proved that domain wall fermions preserve chiral symmetry 
properties very well\cite{Blum}. Presently, these Domain 
wall fermions are being used to study the QCD phase transition\cite{Norman}.
Preliminary tests seem to  indicate the presence of zero modes that
break the anomalous $U_A(1)$ symmetry even above the phase transition.
This suggests that it should be possible to extend the above discussion 
of the critical behavior of the topological charge to such a study. This 
would give a more quantitative grasp of the two flavor QCD phase transition 
than has been possible before.

The above solutions to the Ginsparg-Wilson relation obey an extra 
hermiticity property, namely $D^\dagger = \gamma_5 D \gamma_5$. Based on 
this it is possible to write a formula for the chiral condensate in terms 
of the eigenvalues of the Dirac operator. It is easy to show that
$aD - 1$ is unitary whose eigenvalues can be represented as 
$-\exp(i\theta),\;\; -\pi\leq \theta \leq \pi$. The exact 
zero modes of $D$ correspond to eigenvalues at $\theta = 0$. Thus
\begin{eqnarray}
\lefteqn{\langle \bar\Psi[1- \frac{1}{2}aD]\Psi \rangle \;\;=\;\;}
\nonumber \\
&&\;\;\;\;\;\;\frac{1}{a^3}\int_{-\pi}^\pi d\theta\;\rho(\theta) 
\frac{1 + \cos(\theta) + i\sin(\theta)}
{2(1 - \cos(\theta) - i\sin(\theta) + am)},
\end{eqnarray}
where $\rho(\theta)$ is the normalized density of eigenvalues
$\int d\theta\;\rho(\theta) = 4N_f N_c$. Due to the hermiticity of $D$ the 
eigenvalues come in complex conjugate pairs, so that 
$\rho(\theta)$ is a symmetric function. In the $m\rightarrow 0$ 
limit it is easy to show that
\begin{equation}
\langle \bar\Psi[1- \frac{1}{2}aD]\Psi \rangle = 
\frac{1}{a^3}\pi\rho(0).
\end{equation}
This is the Banks Casher formula\cite{Banks}. In the present context
it is easy to see that ${\rm tr}(\gamma_5 D)$ gets contribution only
from the subspace of eigenvectors of $D$ with eigenvalues $2/a$. 
However this contribution is also related to $(n_+-n_-)$ due to
the fact that ${\rm tr}(\gamma_5) = 0$. This is one way to understand 
the origin of eq.(\ref{eq:anom}).

Results from simulations using the Dirac operator suggested 
by Neuberger in the Schwinger model\cite{Sch} and the domain wall fermions 
in quenched QCD\cite{Pavlos} give clear evidence for divergences in the
quenched chiral condensate due to exact zero modes on finite lattices.
Such singularities are expected due to the anomaly as discussed above, but 
have not been seen in realistic simulations involving staggered fermions. 
This perhaps reflects the fact that the breaking of chiral symmetry in 
the quenched theory, is represented well by Ginsparg-Wilson fermions even 
at finite lattice spacings. Thus, quenched lattice QCD with Ginsparg-Wilson 
fermions may give a more definitive answer to the question about the 
reliability of the quenched approximation.

In conclusion, we have shown how Ginsparg-Wilson fermions
reproduce the low energy physics of QCD at finite lattice
spacings. The exact chiral symmetry of the action can 
in principle be used to derive a host of soft pion theorems
even at finite lattice spacings. We have concentrated on the 
Goldstone's theorem and obtained interesting consequences from 
a solution to the $U(1)$ problem. We have also suggested how
the critical behavior of the topological susceptibility can 
confirm the second order nature of the QCD phase transition
with two massless quarks.
If simulations involving Ginsparg-Wilson fermions are feasible it is 
very likely that the results would reflect physical reality more 
easily than the conventional fermion formulations.

I thank Tanmoy Bhattacharya, Norman Christ, Maarten Golterman and 
Rajamani Narayanan for helpful discussions. I also thank Robert 
Mawhinney and Shigemi Ohta for organizing a stimulating workshop 
on {\it Fermion Frontiers in Vector Lattice Gauge theories} at the 
Brookhaven National Laboratory where I recognized the possible 
usefulness of the above ideas to the physics community. I thank 
Peter Hasenfratz for pointing me to the history of the Ginsparg-Wilson 
relation and reviving our interest towards it. Finally I wish to thank 
Ferenc Niedermayer and Wolfgang Bietenholz for their comments on 
typograghical errors.

\end{document}